\documentclass[aps,amsmath,amssymb,reprint]{revtex4-1}
\usepackage{graphicx}
\usepackage{dcolumn}
\usepackage{bm}
\usepackage[english]{babel}
\newcommand{\p}{$\%$}
\newcommand{\pat}{{${~at.}$}\%}
\newcommand{\muB}{$\mathrm{\mu_{B}}$}

\newcommand{\Ts}{${T_{s}}$}
\newcommand{\FeC}{$\mathrm{Fe_{0.8}C_{0.2}}$}
\newcommand{\tifc}{$\mathrm{Fe_{3}C}$}
\newcommand{\tefc}{$\mathrm{Fe_{4}C}$}
\newcommand{\bhf}{$\mathrm{B_{hf}}$}
\newcommand{\Hf}{$\Delta$H$_{f}^{\circ}$}
\newcommand{\ife}{$\mathrm{^{57}Fe}$}

\begin{document}
\title{Structural and magnetic properties of co-sputtered Fe$_{0.8}$C$_{0.2}$ thin films}
\author{Prabhat Kumar}\address{UGC-DAE
Consortium for Scientific Research, University Campus, Khandwa
Road, Indore 452 001, India}
\author{V. R. Reddy}\address{UGC-DAE
Consortium for Scientific Research, University Campus, Khandwa
Road, Indore 452 001, India}
\author{V. Ganesan}\address{UGC-DAE
Consortium for Scientific Research, University Campus, Khandwa
Road, Indore 452 001, India}
\author{I. Sergueev, O. Leupold and H.-C. Wille}
\address{Deutsches Elektronen-Synchrotron DESY, Notkestrasse 85, D-22607 Hamburg, Germany}

\author{Mukul Gupta} \email {mgupta@csr.res.in}
\address{UGC-DAE Consortium for Scientific Research, University Campus, Khandwa
Road, Indore 452 001, India}

\date{\today}


\begin{abstract}
We studied the structural and magnetic properties of \FeC~thin
films deposited by co-sputtering of Fe and C targets in a direct
current magnetron sputtering (dcMS) process at a substrate
temperature (\Ts) of 300, 523 and 773\,K. The structure and
morphology was measured using x-ray diffraction (XRD), x-ray
absorption near edge spectroscopy (XANES) at Fe $L$ and C
$K$-edges and atomic/magnetic force microscopy (AFM, MFM),
respectively. An ultrathin (3\,nm) $^{57}$\FeC~layer, placed
between relatively thick \FeC~layers was used to estimate Fe
self-diffusion taking place during growth at different \Ts~using
depth profiling measurements. Such $^{57}$\FeC~layer was also used
for $^{57}$Fe conversion electron M\"{o}ssbauer spectroscopy
(CEMS) and nuclear resonance scattering (NRS) measurements,
yielding the magnetic structure of this ultrathin layer. We found
from XRD measurements that the structure formed at low
\Ts~(300\,K) is analogous to Fe-based amorphous alloy and at high
\Ts~(773\,K), pre-dominantly a \tifc~phase has been formed.
Interestingly, at an intermediate \Ts~(523\,K), a clear presence
of \tefc~(along with \tifc~and Fe) can be seen from the NRS
spectra. The microstructure obtained from AFM images was found to
be in agreement with XRD results. MFM images also agrees well with
NRS results as the presence of multi-magnetic components can be
clearly seen in the sample grown at \Ts~= 523\,K. The information
about the hybridization between Fe and C, obtained from Fe $L$ and
C $K$-edges XANES also supports the results obtained from other
measurements. In essence, from this work, experimental realization
of \tefc~has been demonstrated. It can be anticipated that by
further fine-tuning the deposition conditions, even single phase
\tefc~phase can be realized which hitherto remains an experimental
challenge.
\end{abstract}

\maketitle

\section{Introduction}
\label{1} Tetra iron compounds (Fe$_{4}$X) are an interesting
class of compounds known to exhibit higher (than Fe) magnetic
moment (\emph{M}),~\cite{MATAR1988EDP, BLANCA2009PSS,
TAKAHASHI2011JMMM, FANG2014RSC} and spin-polarization ratio
(SPR)~\cite{KOKADO2006PRB, LV2013JMMM}. In addition, metallic
resistivity and corrosion resistance of Fe$_{4}$X make them
attractive for applications. The structure of Fe$_{4}$X compounds
have been classified as anti-perovskite [P\=43m (215)] and
metalloid element such as X = B, C, N occupy the body center
position within the face center cubic (fcc) lattice of host Fe.
Theoretically predicted values of \emph{M} for Fe$_{4}$B,
Fe$_{4}$C and Fe$_{4}$N are 2.57, 2.42 and
2.62\,$\mu$$_{B}$~\cite{MATAR1988EDP, BLANCA2009PSS,
TAKAHASHI2011JMMM, LV2013JMMM, FANG2014RSC}, respectively. Such an
enhancement in \emph{M} occurs due to volume expansion of Fe
lattice. An expansion in volume produces higher density of states
near the fermi level due to contraction of \emph{d}-band (relative
to Fe) and therefore results in higher \emph{M}. The increase in
\emph{M} due to volume expansion is known as magneto-volume
effect~\cite{HOUARI2010JMMM}. Experimentally, \emph{M} of
Fe$_{4}$N has been achieved close to its theoretical values
($\approx$2.4\,$\mu$$_{B}$) in several
works~\cite{2019_arxiv_Fe4N,APL11_Ito_Fe4N,TAKAHASHI2001RSC,KRAMER2004PRB}.
Such enhancement in \emph{M} with respect to pure Fe (\emph{M} of
Fe = 2.2\,\muB) make them very useful in various applications e.g.
spintronics, magnetic data storage devices~\cite{KOHMOTO1991IEEE},
permanent magnets, spin-injection electrodes~\cite{KOKADO2006PRB},
etc. In addition, recent density functional theory calculation
suggests that the SPR for Fe$_{4}$X (X = B, C, N) comes out to be
84, 88 and 61\p, respectively~\cite{LV2013JMMM}.

Furthermore, \tefc~is thermally more stable than others (e.g.
Fe$_{4}$N or Fe$_{4}$B) due to smaller Fe--C bond
distance~\cite{FANG2011PRB, FANG2012PRB, ZHANG2012JMMM,
LV2013JMMM, FANG2014RSC}. Theoretical calculation suggests that
Fe--C bond length at 1.8739\,\AA~is smaller as compared to Fe--B
(1.9027\,\AA) or Fe--N (1.8899\,{\AA})~\cite{LV2013JMMM}.
Likewise, other transition metal carbides also show higher thermal
stability than nitrides of similar composition. Their thermal
stability can also be expressed in terms of Debye-temperature
($\theta_B$), which is a classical limit of quantum model of
Einstein-Debye specific heat for solids. Guillermit \emph{et
al.}~\cite{GUILLERMET1989PRB, GUILLERMET1992PRB} calculated
$\theta_B$ for some transition metal carbides and nitrides and
found that in general, $\theta_B$ of carbides are higher than that
of nitrides.

In spite of very interesting magnetic properties, higher SPR and
thermal stability, experimental reports on \tefc~are almost
non-existent and single phase \tefc~has not yet been synthesized.
Recently, interatomic potential for Fe-C systems were calculated
using modified embedded atom method (MEAM) and it was predicted
that the presence of C lowers the body center cubic (bcc) to face
center cubic (fcc) transformation barrier~\cite{NGUYEN2018MT,
NGUYEN2018CMS}. Also the activation energy for C diffusion in Fe
was calculated and found to be very small at about
0.8\,eV~\cite{KUAN2018JPCC}. Both, the reduction in barrier for
bcc to fcc and low activation energy point towards favorable
conditions for \tefc~phase formation. However, the enthalpy of
formation energy (\Hf) for \tefc~is positive (about
0.1\,eV~\cite{LV2013JMMM, LIU2016SR}). It may be noted that
\Hf~for early transition metal nitrides and carbides is quite less
e.g. \Hf~= -1.9\,eV for TiC and -3.5\,eV for
TiN~\cite{GUILLERMET1989PRB}. Therefore, synthesis of these
compounds take place straight away using standard methods. Even
with a slight negative \Hf~= -0.15\,eV~\cite{LV2013JMMM,
FANG2014RSC}, single phase Fe$_{4}$N has been reported in several
works~\cite{BLANCA2009PSS}. But for \tefc, a positive value of
\Hf~leads to a thermodynamically unfavorable condition that can
not be achieved utilizing equilibrium processes e.g.
Fischer-Tropsch synthesis (FTS). Recently, FTS was used to
synthesize iron carbide (Fe-C) compounds and resulting phases were
identified as $\theta$-Fe$_{3}$C and $\chi$-Fe$_{5}$C$_{2}$ from
x-ray diffraction and M\"{o}ssbauer~spectroscopy
measurements~\cite{LIU2016SR}.

However, using non-equilibrium processes that are typically found
in physical vapor depositions (PVD) methods, formation of
\tefc~may be feasible. It may be noted that \Hf~of
Fe$_{16}$N$_{2}$ is also positive at about
0.88\,eV/atom)~\cite{TESSIER20004SSS}. Using various thin film
deposition techniques like e-beam evaporation, molecular beam
epitaxy and sputtering, (partial) formation of Fe$_{16}$N$_{2}$
has been reported~\cite{SUGITA1996JAP, TAKAHASHI1994JAP}.
Therefore, formation of \tefc~may also take place using such
non-equilibrium processes. Among PVD methods, co-sputtering is
most convenient and effective to synthesize Fe-C compounds as: (i)
the control of composition can be made precisely by varying the
power of each sputter source (ii) in a confocal geometry,
sputtered species from each source mix before depositing on to a
substrate (iii) ad-atom energies in sputtering are much higher
(typically 10\,eV) than that of thermal evaporation
(0.1-0.2\,eV)~\cite{OHRING2002ACADEMICPRESS}. Thus,
non-equilibrium nature and higher energetics of the depositing
species in sputtering have risen a possibility to prepare \tefc .

Some attempts have been made to synthesize Fe-C thin films by
sputtering of  a compound [Fe+C] target~\cite{Tajima1993JMS,
BABONNEAU1999APL, BABONNEAU2000JAP, MI2004JPCM, Weck2012JMS,
MI2005JAP}, or sputtering of Fe using a mixture of
Ar+CH$_4$~\cite{Jouanny2010JMR} or
Ar+C$_{2}$H$_{2}$~\cite{KIRILYUK1996PRB}. Utilization of compound
target may not be useful to control the composition of Fe-C films
and using gases like CH$_4$ and C$_{2}$H$_{2}$ may also result in
formation of undesired C-H bonds. On the other hand, co-sputtering
of Fe and C from two different sources is a clean method providing
a wide range of option to control the composition by changing the
flux of each sources as demonstrated recently by Furlan et
al.~\cite{FURLAN2015JPCM}. A survey of available literature
suggests that most of the Fe-C thin films reported hitherto are
either amorphous or nanocrystalline. For example, Jouanny \emph{et
al.}~\cite{Jouanny2010JMR} prepared Fe-C thin films by sputtering
of Fe target in Ar/CH$_{4}$ gas environment at \Ts~= 373\,K and
identified phases were $\epsilon$-\tifc~at lower CH$_{4}$ gas flow
and amorphous at higher CH$_{4}$ gas flow. Tajima \emph{et al.}
deposited Fe-C thin films at 623\,K by rf magnetron sputtering of
a compound target ~\cite{Tajima1993JMS} and phases identified were
$\alpha$-Fe, \tifc~Fe$_{5}$C$_{2}$. Weck \emph{et
al.}~\cite{Weck2012JMS} deposited Fe-C thin films by ac magnetron
sputtering and reactive cathodic arc evaporation of a compound
target (12 and 16 at.$\%$ C). Resulting films were found to be
nano-crystalline bcc-Fe. Subsequently, Babonneau \emph{et
al.}~\cite{BABONNEAU2000JAP} deposited Fe$_{1-x}$C$_{x}$
(0.26$\leq$~x~$\leq$0.74) thin films at various substrate
temperatures but resulting films were found to be amorphous.
Similarly, Mi \emph{et al}.~\cite{MI2004JPCM} deposited Fe-C thin
films by sputtering of a compound target (at room temperature) and
also found amorphous Fe-C phases. More recently, Furlan \emph{et
al.}~\cite{FURLAN2015JPCM} deposited Fe-C thin films by
co-sputtering of Fe and C from two separate targets (at 300\,K)
and varied C concentration from 20.8 to 71.8\pat. It was observed
that resulting Fe-C films were amorphous, irrespective of the
amount of C in Fe.

In this work, we synthesized Fe-C thin films by co-sputtering of
Fe and C at different substrate temperature (\Ts) with a nominal
composition of \FeC~so as to attempt the first ever experimental
realization of \tefc. Samples were characterized using x-ray
diffraction (XRD) for their structure and the surface and magnetic
morphology has been obtained from atomic and magnetic force
microscopy (AFM, MFM) measurements, respectively. The information
about the nature of bonding has been obtained from XANES
measurements at C $K$ and Fe $L$-edges. The magnetization of
samples were studied using conversion electron M\"{o}ssbauer
spectroscopy (CEMS) and synchrotron based nuclear resonant
scattering (NRS). In addition, from secondary ion mass
spectroscopy (SIMS) depth-profiles measurements, Fe self-diffusion
was estimated and compared with a pure Fe sample prepared under
similar conditions. We found that at low \Ts~the addition of C
suppresses Fe diffusion but at high \Ts~it augments. From above
mentioned measurements, we found that the room temperature grown
\FeC~film was amorphous but at high \Ts, phases formed are
crystalline. At the highest \Ts~(773\,K), pre-dominantly a
\tifc~phase has been formed but at an intermediate \Ts~of 523\,K,
a clear presence of \tefc~(along with \tifc~and Fe) can be seen.
Observed results demonstrate the possibilities for formation of
crystalline Fe-C phases by co-sputtering. The structural and
magnetic properties of thus formed Fe-C phases are presented and
discussed in this work.

\section{Experimental Details}
\label{2} Fe and C targets were co-sputtered to prepare
iron-carbon (Fe-C) thin films. They were deposited at substrate
temperature (\Ts) = 300, 523 and 773\,K with nominal
stoichiometric composition for \tefc~or \FeC. Their layer
structure was:
{C(5\,nm)$\vert$$\mathrm{^{natural}}$Fe-C(70\,nm)$\vert$$\mathrm{^{57}}$Fe-C(3\,nm)$\vert$$\mathrm{^{natural}}$Fe-C(100\,nm)$\vert$C(10\,nm)$\vert$sub.(Si/Quartz).
The C concentration was evaluated using: (i) number of monolayer
(n) per unit area per unit mole: n = mass-number/mass-density (ii)
thicknesses of Fe and C equivalent to stoichiometric \FeC~is
n$\mathrm{_{Fe}}$$\times$0.8 and n$\mathrm{_{C}}$$\times$0.2 (iii)
deposition rates for Fe and C has been optimized to achieve
thicknesses obtained for \FeC~composition.

The chamber was evacuated down to a base pressure of
2$\times$10$^{-7}$\,hPa. Sputtering was carried out using pure Ar
(purity 99.9995\p) gas at a working pressure of
3$\times$10$^{-3}$\,hPa due to gas flow of 20\,sccm. In this layer
structure, the $\mathrm{^{natural}}$Fe-C (hereafter Fe-C) was
prepared by co-sputtering of a $\phi$3\,inch Fe (purity 99.95\%)
and $\phi$3\,inch C (purity 99.999\%) target at the sputtering
power of 100 and 156\,W, respectively. On the other hand, for the
$^{57}$Fe-C, the same C target and a $\phi$1\,inch $^{57}$Fe
(enrichment 95.0\%, purity 99.95\%) target was used and their
sputtering powers were kept at 70 and 11\,W, respectively. To
prevent Si diffusion from the substrate and surface contamination
of samples, a buffer and capping layer of C was always deposited
at room temperature prior to or after the deposition of \FeC~film.
For better uniformity of the films, substrate holder was rotated
at 60\,rpm at a distance of about 12\,cm from the target. The
structural growth of samples was characterized by XRD measurements
using Bruker D8 Advance diffractometer equipped with Cu
\emph{K$\alpha$} x-ray source. The surface morphology and magnetic
domain growth of samples was obtained from AFM and MFM
measurements (in tapping mode), respectively. During MFM
measurements, the cantilever has been lifted up at a height of
50\,nm from the surface, to minimize the effect of topography. The
electronic structure of the deposited samples have been
investigated using Fe $L$-edges and C $K$-edge XANES measurements
at BL-01 beamline~\cite{INDUS2BL01}, Indus-2 synchrotron radiation
source, RRCAT, Indore, India in total electron yield (TEY) mode
under UHV conditions. The local structural and magnetic properties
have been determined using CEMS and grazing incidence NRS
measurements. The angle of incidence during NRS measurements was
kept fixed at the critical angle ($\theta_c$) $\approx$
0.21$^{\circ}$ and they were performed at P01
beamline~\cite{PETRA3} of DESY, Petra III synchrotron radiation
source at Hamburg, Germany.

In addition, we also deposited reference samples: (i) pure C thin
films at \Ts~= 300, 523 and 773\,K and, (ii) pure Fe thin film
with a $^{57}$Fe marker layer. The process parameters during
growth of these samples were kept similar to those described for
\FeC~samples. The structure of pure Fe samples was kept as:
$\mathrm{^{natural}}$Fe(70\,nm)$\vert$$\mathrm{^{57}}$Fe(3\,nm)$\vert$$\mathrm{^{natural}}$Fe(100\,nm)$\vert$
sub.(Si/Quartz). They were grown at \Ts~= 300, 523, 648 and
773\,K. In this case, Fe and $^{57}$Fe were sputtered
alternatively from two different sources (at 100\,W) and the
$^{57}$Fe target was prepared by adding $^{57}$Fe foils within the
race track area of $\mathrm{^{natural}}$Fe target. SIMS depth
profiles from \FeC~and Fe samples have been compared to deduce the
effect of C addition on Fe self-diffusion at various \Ts.

\section{Results and Discussion}
\label{3}
\subsection{Structure and microstructure}\label{3.1}

Figure~\ref{xrd} shows XRD pattern of a pure Fe film (reference
sample) and \FeC~thin films deposited at \Ts~= 300, 523 and
773\,K. The XRD pattern (fig.~\ref{xrd}a) of Fe thin film
deposited at \Ts~= 300\,K shows growth of the film oriented along
(200) plane and when the \Ts~exceeds 523\,K, the (110) plane
becomes prominent. Generally, such variations in preferred
orientations are not unexpected and have been observed in several
cases like TiN~\cite{SUNDGREN1985TSF, OH1993JAP, PATSALAS2000SCT},
AlN~\cite{MEDJANI2006TSF} etc. and have been explained considering
alterations in the adatom mobility, stress/strain, and surface
energy due to enhanced \Ts~\cite{PATSALAS2000SCT, MEDJANI2006TSF}.
As shown later in this work (section~\ref{3.4}), the Fe
self-diffusion does not increase appreciably on increasing the
\Ts~(in fact it decreases slightly at 523\,K), therefore arguments
related to adatom mobility may not be valid for the change
observed in the preferred orientation of Fe films. In a recent
work, Sch\"{o}necker \emph{et al.} calculated the thermal surface
free energy and stress of iron at different temperatures and found
that surface stress for the (001) surface was much smaller than
that of (110) surface at low temperature but at high temperatures
they become similar~\cite{Schönecker2015SR}. The changes observed
in the preferred orientation in our Fe films can be understood
from this argument. On the other hand, the width of XRD peaks
become narrow as \Ts~increases. The crystallite size (t) has been
calculated using Debye Scherrer formula, t =
0.96$\lambda$/$\beta$cos $\theta$, where $\lambda$ is wavelength
of the x-rays, $\beta$ is angular full width half maxima of the
Bragg reflection centered at 2$\theta$. By increasing the \Ts~from
300 to 523 and 773\,K, t increases from 18$\pm$0.5 to 28$\pm$0.5
and 46$\pm$2.0\,nm, respectively.

The XRD pattern (fig.~\ref{xrd}b) of the \FeC~thin film deposited
at \Ts~= 300\,K shows a broad reflection centered around 2$\theta$
= (44.58$\pm$0.1)$^{\circ}$ signifying that it has attained an
amorphous structure. In an amorphous system, the average nearest
neighbor distance (d) can be calculated using : d =
1.23$\lambda$/2 sin $\theta$, where $\theta$ is center of the
broad reflection and 1.23 is a geometric factor which rationalizes
the nearest neighbor distance with the spacing between,
``pseudo-close packed planes"~\cite{KARTZ1964}. From here, we get
d = 2.5\,{\AA}, a value typically found in iron-based amorphous
alloys.~\cite{MCHENRY1999}. On the other hand, samples deposited
at higher \Ts~ show a number of peaks. Observed peak positions for
the sample deposited at \Ts~= 523 and 773\,K are similar and their
intensity is increasing with \Ts. These peak positions match-well
with that of $\mathrm{\theta}$-Fe$\mathrm{_{3}}$C (PDF\#89-2867)
and $\alpha$-Fe (PDF\#870721) phases. A general observation shows
that by increasing \Ts,~peak broadening decreases indicating
increase in crystallite size. The crystallite size of the evolved
phases in the sample deposited at \Ts~= 523\,K~ comes out to be
17$\pm$1\,nm of $\theta$-Fe$_{3}$C phase corresponding to (210)
reflection centered at (43.74$\pm$0.01)\,$^{\circ}$ and
26$\pm$1~\,nm for $\alpha$-Fe phase corresponding to (110)
reflection centered at (44.75$\pm$0.01)\,$^{\circ}$. Similarly,
crystallite size is 42$\pm$2 and 28$\pm$2 nm for
$\theta$-Fe$_{3}$C and $\alpha$-Fe phases, respectively for the
sample deposited at \Ts~= 773\,K.

\begin{figure} \center
\vspace{-12mm}
\includegraphics [width=90mm,height=130mm] {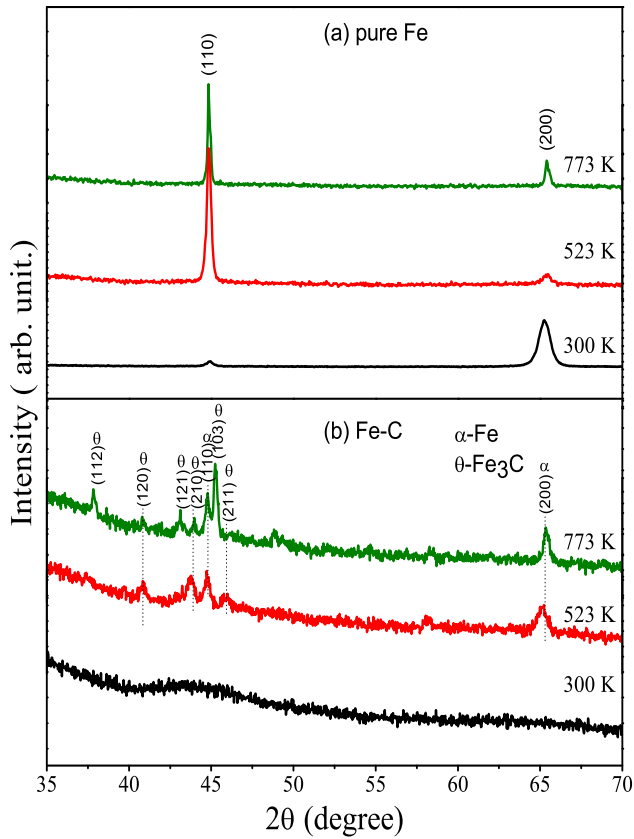} \vspace{2mm}
\caption{\label{xrd} (Color online) X-ray diffraction pattern of
(a) pure Fe and (b) \FeC~thin films deposited at various substrate
temperature.} \vspace{-5mm}
\end{figure}

The solubility of C in bcc Fe under ambient condition is very
small ($\approx$0.2 at\%)~\cite{Callister}. Beyond this limit, the
presence of C produces a disordered crystal structure due to
immiscibility of C with Fe. This results in formation of an
amorphous phase in our sample and show agreement with previous
reports~\cite{FURLAN2015JPCM}. Increasing substrate temperature
pushes C at interstitial position of the orthorhombic crystal
lattice of Fe, this results in growth of
$\mathrm{\theta}$-Fe$\mathrm{_{3}}$C phase. However, peaks
corresponding to un-reacted $\alpha$-Fe are also visible.

Figure~\ref{spm} shows surface and magnetic morphology of the
\FeC~ thin films deposited at various~\Ts. Images were processed
using WSxM software package~\cite{WSXM2007}. For better
understanding of grain and domain growth, we have plotted height
and frequency distribution profile of the AFM
(fig.~\ref{profile}(a)) and MFM (fig.~\ref{profile}(b)) images,
respectively. In the AFM image of the sample deposited at \Ts~=
300\,K (fig.~\ref{spm}(a)), the grains are very small. It is also
clear from the height distribution (H$_{d}$) profile
(fig.~\ref{profile}(a)). On the other hand, the AFM image of the
sample deposited at \Ts~= 523\,K shows enhancement in the grains
(fig.~\ref{spm}(b)) and their H$_{d}$ profile shows that they are
nearly equal in size. The sample deposited at \Ts~= 773\,K have
much larger grains compared to other two samples
(fig.~\ref{spm}(c)). The H$_{d}$ profile of the AFM image shows
uniform growth of larger grain along with smaller grains. This
shows that increase in \Ts~results in formation of larger grains.
This is in agreement with the XRD results as the grains of
$\mathrm{\theta}$-Fe$\mathrm{_{3}}$C become larger with increase
in \Ts.

The MFM images and their frequency distribution profile (F$_{d}$)
show that the sample deposited at \Ts~= 300\,K may have very small
magnetic domains (fig.~\ref{spm}(b) and ~\ref{profile}(b)) and
they are not clearly visible. The sample deposited at \Ts~= 523\,K
have large magnetic domains (fig.~\ref{spm}(d)) with a systematic
change in the cantilever frequency (fig.~\ref{profile}(b)). On the
other hand, even larger magnetic domains can be seen in the MFM
image (fig.~\ref{spm}(e)) of the sample deposited at \Ts~= 773\,K.
However, the magnetic domains in MFM image is following the
pattern similar to topographic changes observed in the AFM
(fig.~\ref{spm}(e)). In addition, the frequency distribution
profile (fig.~\ref{profile}(b)) shows a change in the frequency
similar to change in H$_{d}$ profile (fig.~\ref{profile}(a)). This
is an indication of presence of two kinds of magnetic domain, one
with larger and another with smaller size. The MFM image of the
sample deposited at Ts~= 523\,K shows that the structural and
magnetic morphologies are different. The F$_{d}$ profile shows
various frequency maxima with different magnitude
(fig.~\ref{profile}(b)). This change in the magnetic field of the
sample deposited at \Ts~= 523\,K may arise if several magnetic
phases are present together. The presence of such magnetic phases
can be investigated using a local magnetic probe e.g.
M\"{o}ssbauer spectroscopy based techniques like CEMS and NFS.
Results of CEMS and NFS are presented in section \ref{3.3}.

\begin{figure}
    \begin{center}
        \includegraphics[width=40mm,height=35mm]{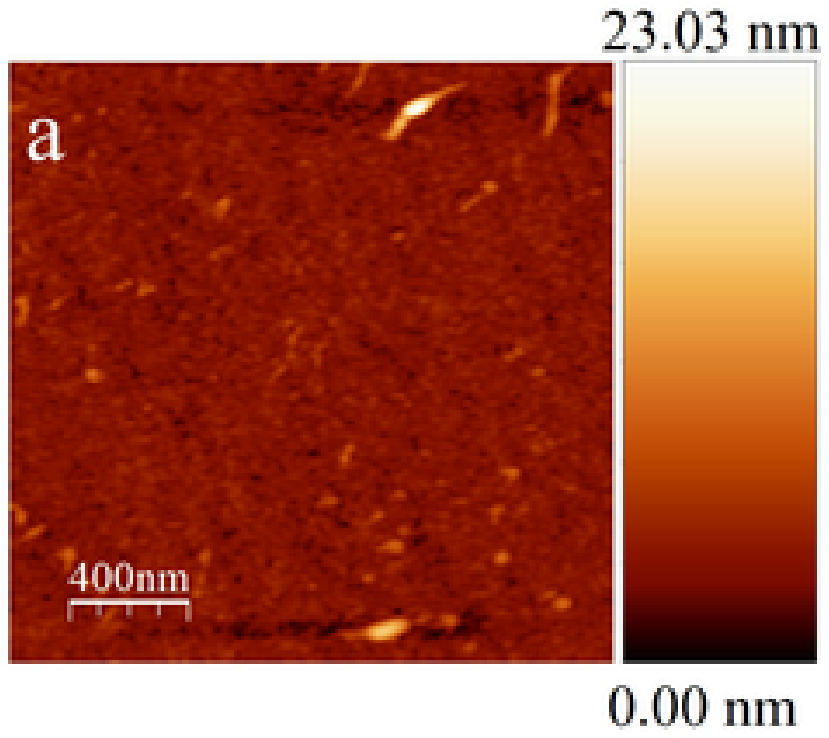}
        \includegraphics[width=40mm,height=35mm]{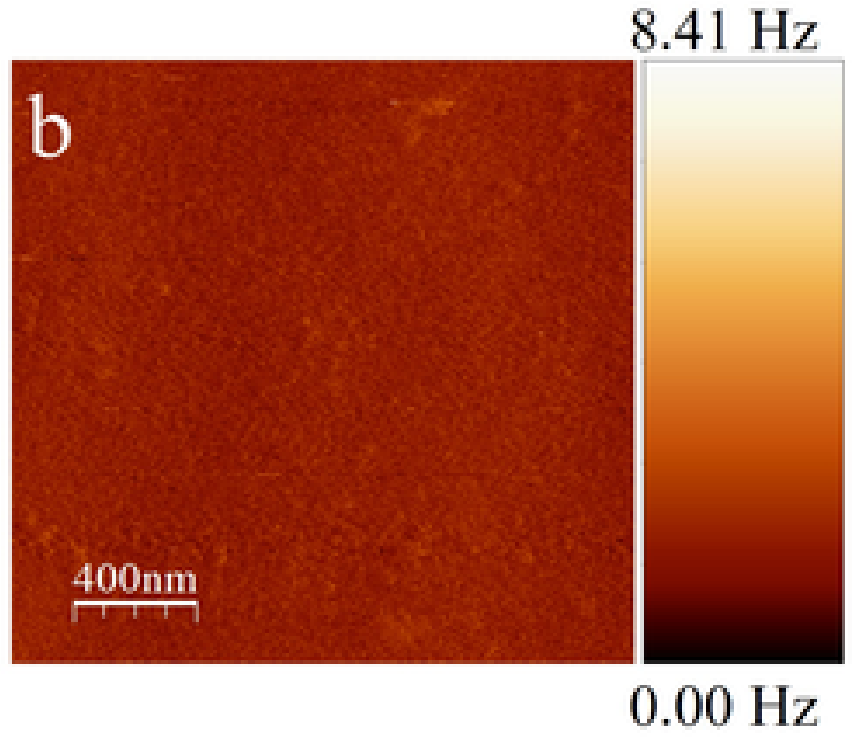}
        \includegraphics[width=40mm,height=35mm]{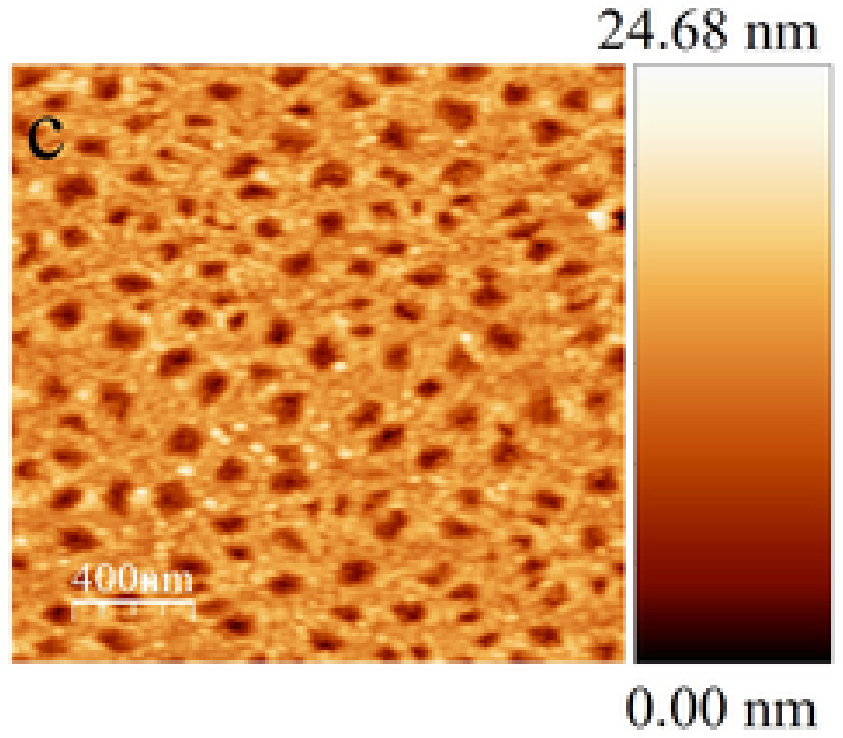}
        \includegraphics[width=40mm,height=35mm]{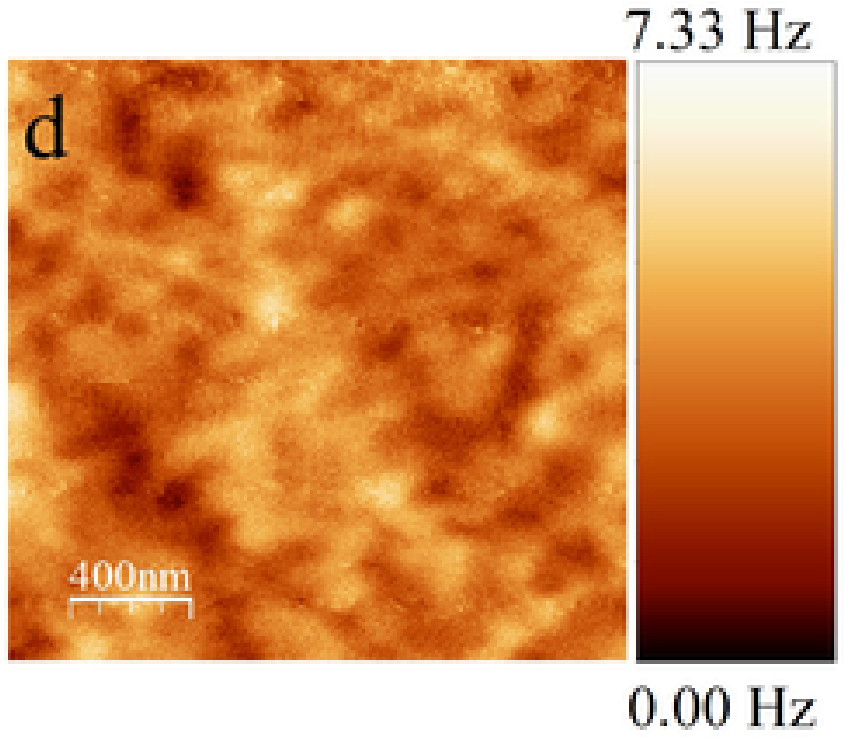}
        \includegraphics[width=40mm,height=35mm]{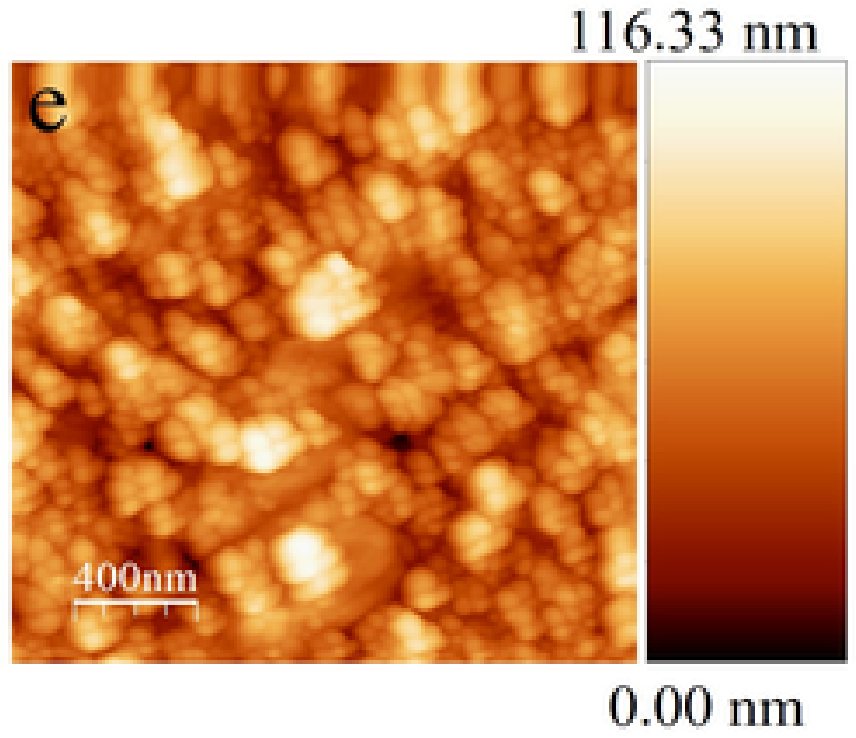}
        \includegraphics[width=40mm,height=35mm]{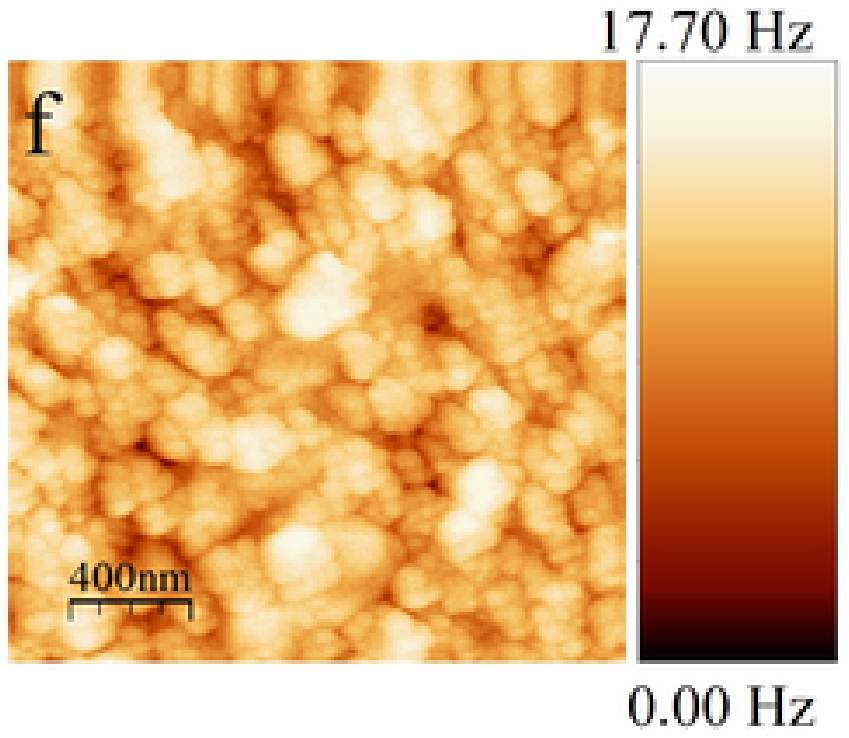}
        \caption{\label{spm} (Color online) Atomic force microscopy (AFM) (a), (c) and (e), magnetic
        force microscopy (MFM) (b), (d) and (f) images of \FeC~samples deposited at 300, 523 and 773\,K
        respectively with color scale. The color scale in AFM images represent height and in MFM images it represent cantilever frequency vibrating along z-axis. The X$\times$Y scale in all the images are 2$\times$2~$\mu$m.}
    \end{center}
\end{figure}

\begin{figure} \center
\vspace{10mm}
\includegraphics[width=80mm,height=65mm]{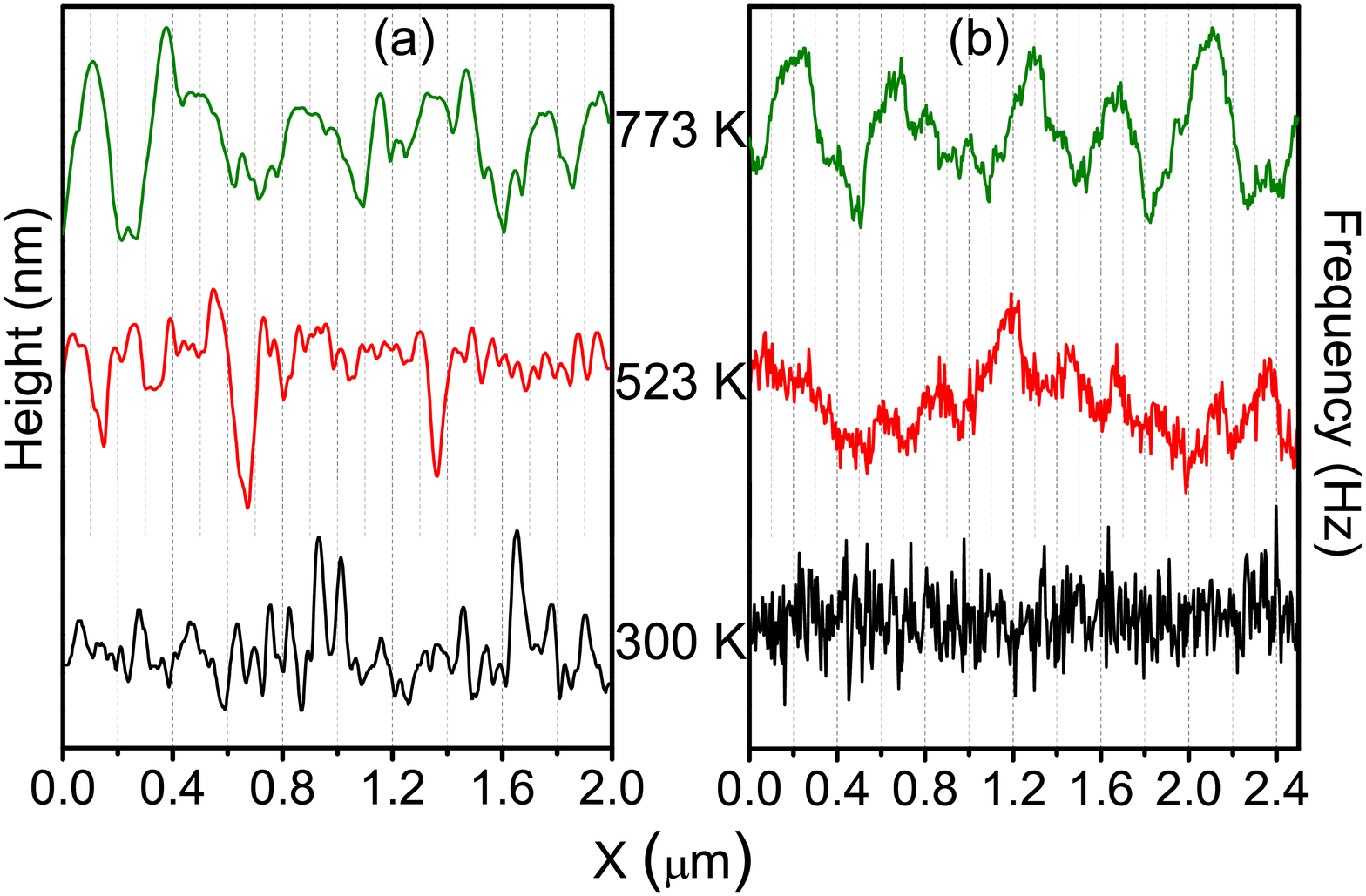}
\caption{\label{profile} (Color online) Height (a) and frequency
(b) distribution profile of atomic force microscopy (AFM) and
magnetic force microscopy (MFM) images of the as-deposited
\FeC~samples respectively. The Profile from AFM is taken along
x-axis. In MFM profile, it is taken along the the diagonal of the
images}
\end{figure}

\subsection{Electronic structure}\label{3.2}

Local electronic structure was probed using synchrotron based
XANES measurements at Fe $L_{3,2}$ and C $K$ absorption edges.
Fig.~\ref{feedge} shows Fe $L_{3,2}$-edge XANES spectra of the
\FeC~thin film deposited at various \Ts, following
$2p$$\rightarrow$$3d$ dipole transition. As indicated, the spectra
consist of two prominent transitions occurring due to spin-orbit
splitting of $2p$ orbital in $L_{3}$ (I) and $L_{2}$ (III)
core-shell separated in energy by about 13\,eV, which is a typical
value for Fe. Each sub-spectrum further split in double sub-peak
due to ligand field splitting marked as II and IV in
fig.~\ref{feedge}. Such splitting has been observed in several
transition metal-carbides~\cite{Magnuson2006PRB, Magnuson2009PRB,
Magnuson2012JPCM, FURLAN2015JPCM}. The spectra show two
characteristic change with \Ts~(i) peak become narrow, and (ii)
total integrated intensity of the feature is decreasing as shown
in the inset of the fig.\ref{feedge}.

Generally, a narrow Fe $L_{3,2}$ absorption lines for pure Fe are
observed~\cite{FURLAN2015JPCM}. But, addition of smaller atom e.g.
C, N, O produces splitting due to crystal field and results in
broadening of the resonance lines~\cite{FURLAN2015JPCM}. Their
intensity is proportional to the unoccupied Fe 3\emph{d} states.
Decreasing intensity indicates decrease in charge transfer with
increasing \Ts, this results in reduced unoccupied Fe 3\emph{d}
state at higher \Ts. Earlier reports on Fe and Cr $2p$ XANES
spectra of amorphous system have shown that, the peak intensity
increases with increase in C concentration~\cite{Magnuson2012JPCM,
FURLAN2015JPCM}. Decrease in the unoccupied state show reduced
carbide contribution at higher \Ts. These variation of the spectra
show, there can be possibility of presence of C in un-hybridized
states at higher \Ts. Further insight about the hybridization have
been obtained from C $K$-edge XANES measurements.

Figure~\ref{cedge} shows C \emph{K}-edge XANES spectra of the Fe-C
thin film deposited at various \Ts. The prominent features are
marked as \emph{a}, \emph{b}, \emph{c} and \emph{d}, and shoulders
of \emph{a} and \emph{c} as $a^{\prime}$ and c$^{\prime}$,
respectively. The feature \emph{a} is related with empty $\pi^{*}$
states, it is a combination of (i) $sp^{2}$ and $sp^{1}$
hybridized C states, and (ii) Fe 3$d$-C 2$p$ hybridized states. In
addition, the feature \emph{c} is also a combination of (i) Fe
3\emph{d}-C 2$p$ hybridized states, and (ii) C-O bonding states.
These two features show opposite trend with \Ts, with increase in
\Ts~the intensity of the feature $a$ increases while the intensity
of the feature $c$ decreases. A comparison of the C \emph{K}-edge
spectra of the \FeC~thin film with $a$-C thin film deposited at
\Ts~= 300, 523 and 773\,K shows that, the feature $c$ is shifted
by 0.4\,eV at lower energy side for \emph{a}-C thin films. This
shows in case of \FeC, the feature $c$ is strongly related with Fe
3\emph{d}-C 2$p$ hybridized states. But decrease in the intensity
of this feature shows reduction in carbide contribution as
observed in the Fe $L_{3,2}$-edge spectra. Consequently, increase
in the intensity of the feature $a$ signifies presence of higher
fraction of $sp^{2}$ hybridized C states at higher \Ts. A shift in
the feature $a$ can be observed at higher energy side by 0.3\,eV
for the sample deposited at \Ts~= 523\,K. This shows that, in
addition with reduced carbide contribution at higher \Ts, the
sample deposited at \Ts~= 523\,K have different local structure
compared to samples deposited at \Ts~= 300 and 773\,K. In
addition, a faint feature \emph{b} is observed at 286.4\,eV,
although origin of this feature is yet not clear~\cite{PK2018DRM}.

\begin{figure} \center
\includegraphics [width=90mm,height=75mm] {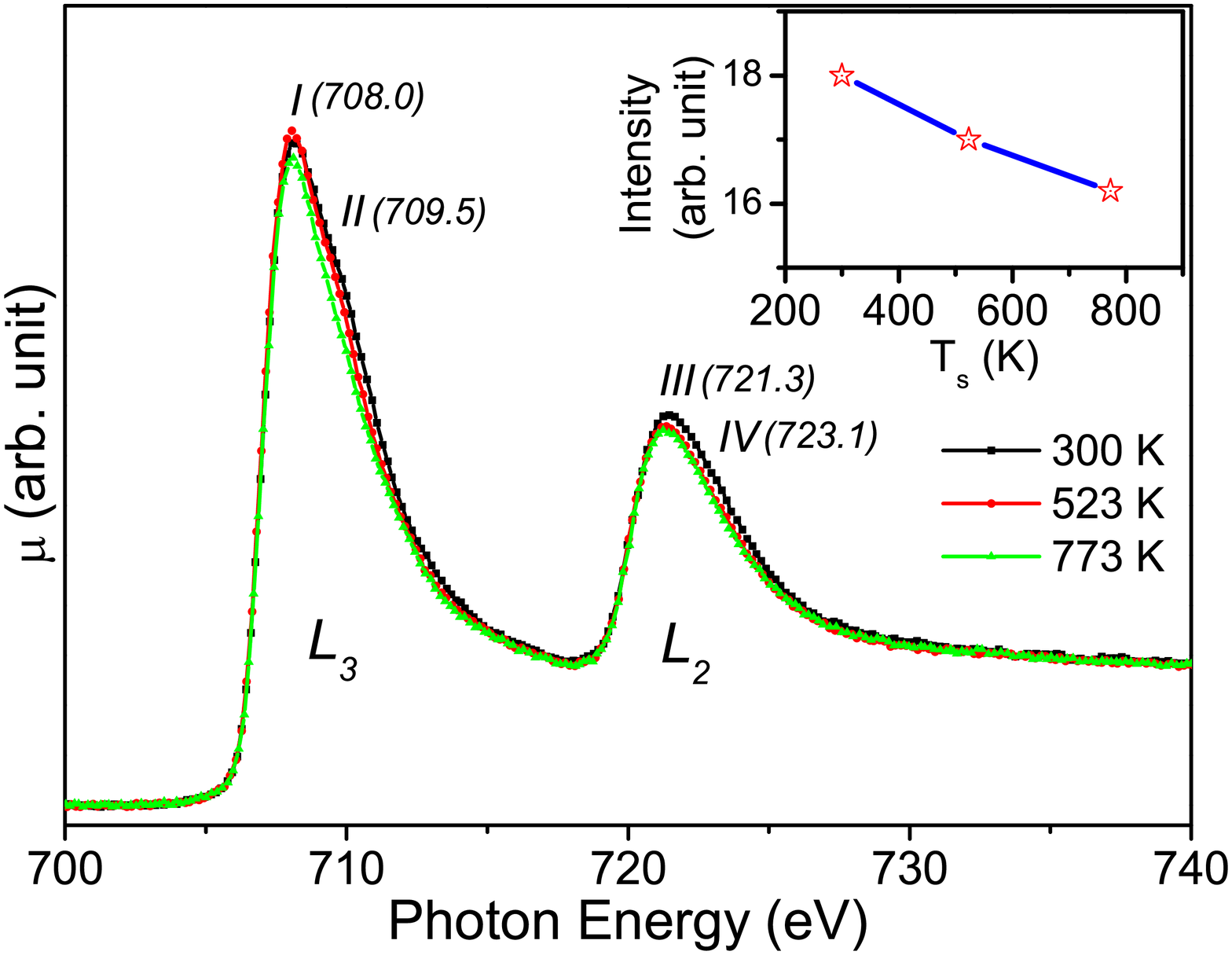} \vspace{-2mm}
\caption{\label{feedge} (Color online) Fe L-edge soft x-ray
absorption spectra of the \FeC~samples deposited  at Ts=300, 523
and 773\,K measured in TEY mode. The inset shows variation of integrated intensity with \Ts~.} 
\end{figure}

\begin{figure} \center
\includegraphics [width=90mm,height=75mm] {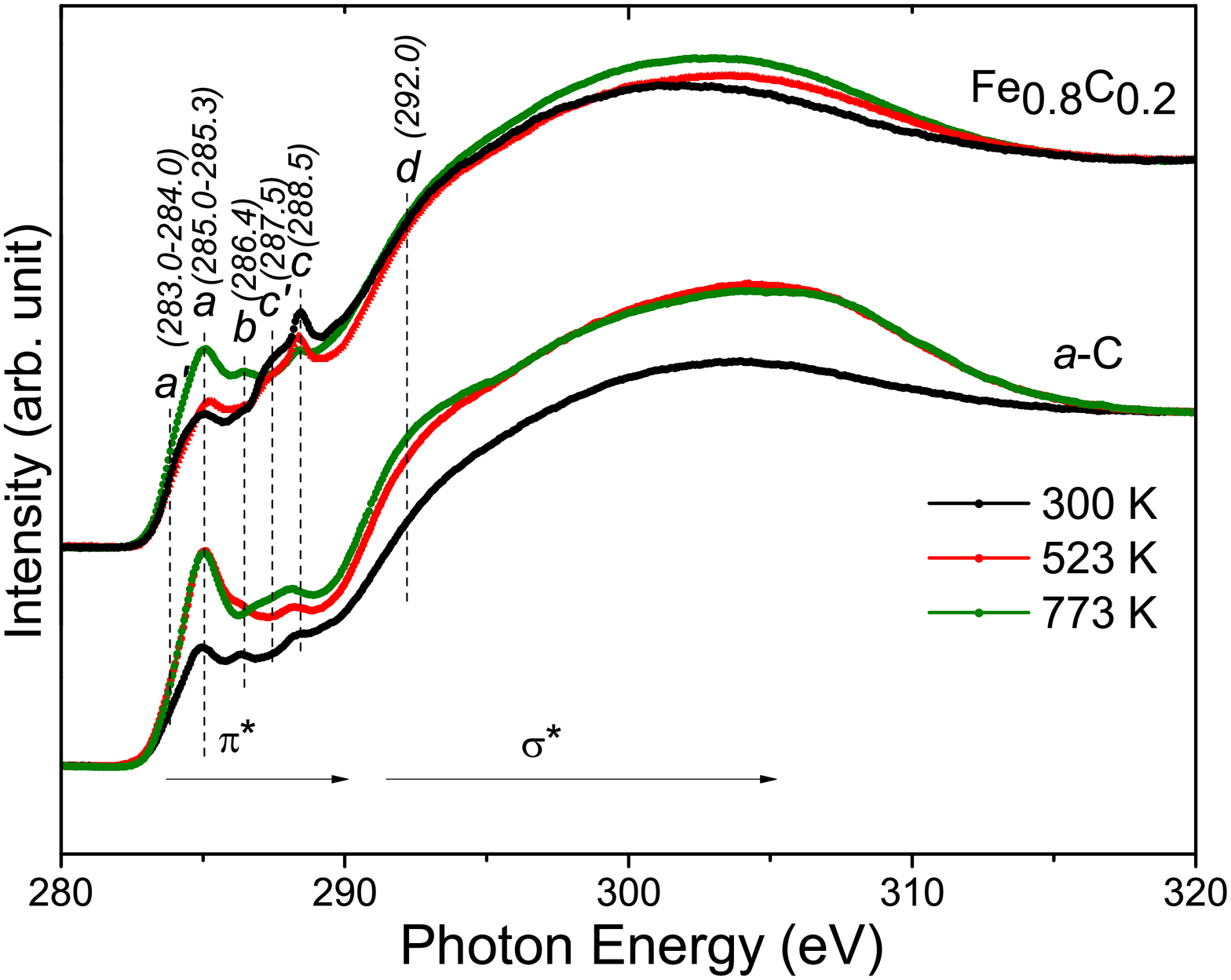} \vspace{-2mm}
\caption{\label{cedge} (Color online) C K-edge soft x-ray
absorption spectra of the \FeC~samples deposited at Ts=300, 523
and 773\,K measured in TEY mode.} 
\end{figure}

In addition, the shoulder $a^{\prime}$ can be solely due to
formation of metal carbide~\cite{FURLAN2015JPCM, Magnuson2009PRB,
Magnuson2006PRB, Magnuson2012JPCM}. The intensity of this shoulder
is faint and does not show any significant change. On the other
hand, the intensity of the shoulder $c^{\prime}$ is decreasing
with increase in \Ts. Another report on nano-crystalline
TiC/$a$-C, shows that the intensity of these features increases
with increasing grain sizes as the carbide contribution increases.
Unlike TiC/$a$-C, decrease in the intensity shows decrease in
carbide contribution with increasing \Ts. This shows presence of
$a$-C in un-hybridized state at higher~\Ts.

\subsection{Magnetic structure}\label{3.3}

Magnetic structure of sandwiched 3\,nm $^{57}$Fe-C layer was
probed using CEMS and NRS measurements. Fig.~\ref{cems} shows
experimental CEMS spectra of samples deposited at various
\Ts~along with fitted data. The CEMS spectrum of the sample
deposited at 300\,K shows broad resonance lines. The broadening of
the resonance lines can be related to lack of long-range ordering
arisen due to immiscibility of C~\cite{MIGLIERINI1997JPCM,
MOSS2011, MIGLIERINI2017PAC}. This spectrum was fitted using one
sextet with hyperfine field (\bhf) of 25.65$\pm$0.34\,T. A \bhf~=
26-30\,T depending on C content can be observed in amorphous Fe-C
alloys~\cite{MIANI1994JALCOM, Amagasa2016HyperfineInter}. This
also in a way confirms the amorphous nature of the sample
deposited at \Ts~= 300\,K. In comparison to this, the spectrum of
the sample deposited at \Ts~= 523\,K shows narrow resonance lines.
It was fitted assuming two sextets S1 and S2 with their \bhf~=
33.20$\pm$0.08 and 20.51$\pm$0.13\,T, respectively. Where, S1 and
S2 can be assigned to $\alpha$-Fe (57\p) and $\theta$-Fe$_{3}$C
(43\p) phases, respectively~\cite{LIU2016SR}.

On the other hand, the CEMS spectrum of the sample deposited at
\Ts~= 773\,K does not show any prominent resonance even after long
counting time (one week). The active \% of M\"{o}ssbauer signal
was already low at about 0.5\p~due to ultrathin $^{57}$Fe-C layer
(3\,nm) but at higher \Ts~(773\,K) the $^{57}$Fe-C layer gets
diffused across the entire depth of sample (shown later from SIMS
depth profiles), active \% of M\"{o}ssbauer signal will reduce
further. It may be noted that the escape depth of electrons in
CEMS is ($\approx $80\,nm)~\cite{FERNANDO2009LACAME} which is
shorter than the total thickness of sample. The large spread of
$^{57}$Fe with \Ts~reduces the effective number of resonating
nuclei leading to poor statistics. Therefore, ensuing magnetic
phases that can be observed in MFM measurements, could not be
resolved from CEMS measurement. Such experimental limitation can
be overcome by doing synchrotron radiation based NRS measurements.

\begin{figure} \center
\includegraphics [width=90mm,height=75mm] {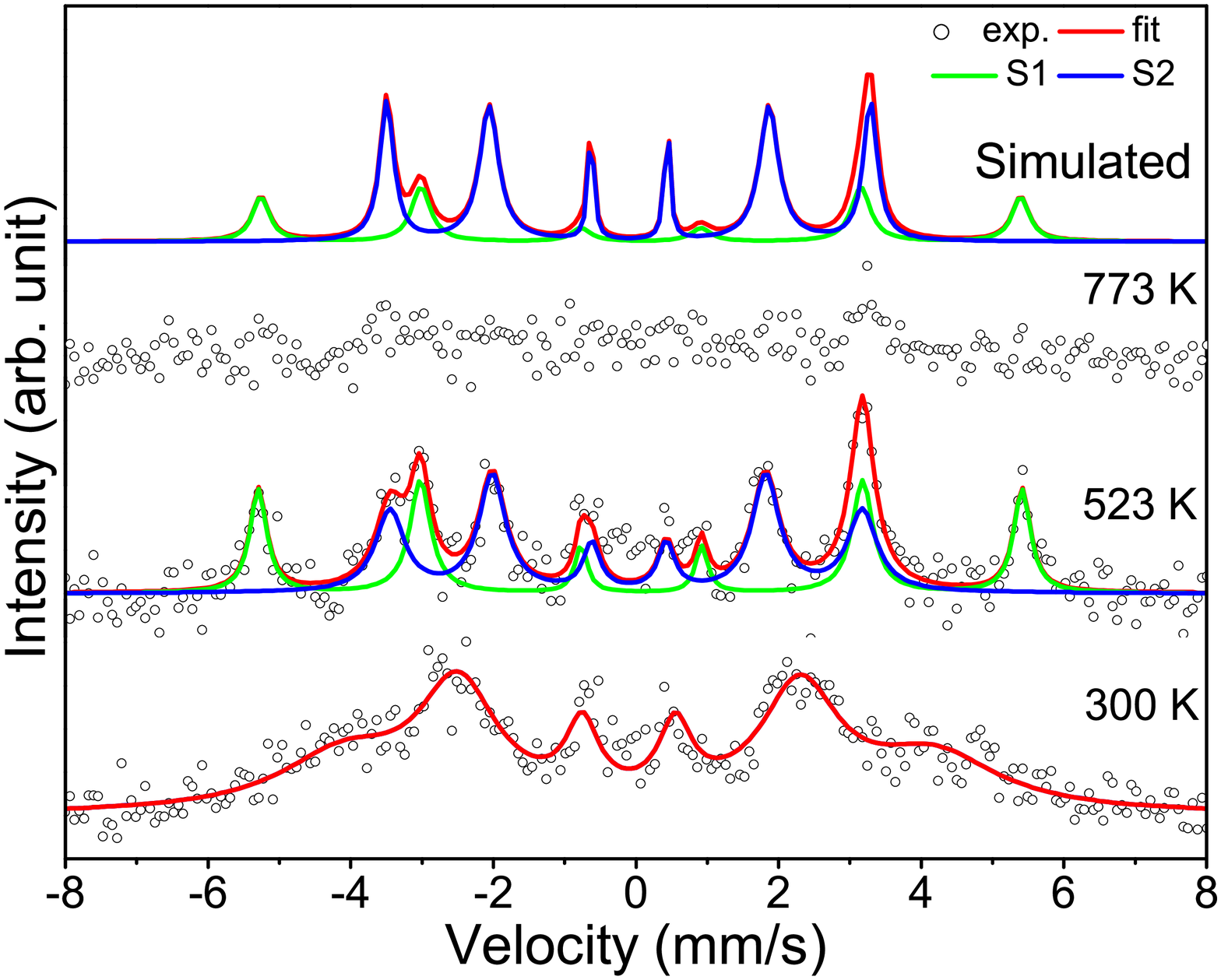} \vspace{-2mm}
\caption{\label{cems} (Color online) Conversion electron
m\"{o}ssbauer spectroscopy of the \FeC~samples grown at \Ts=300,
523 and
773 K with their respective fit.} 
\end{figure}

\begin{figure} \center
\vspace{5mm}
\includegraphics [width=90mm,height=75mm] {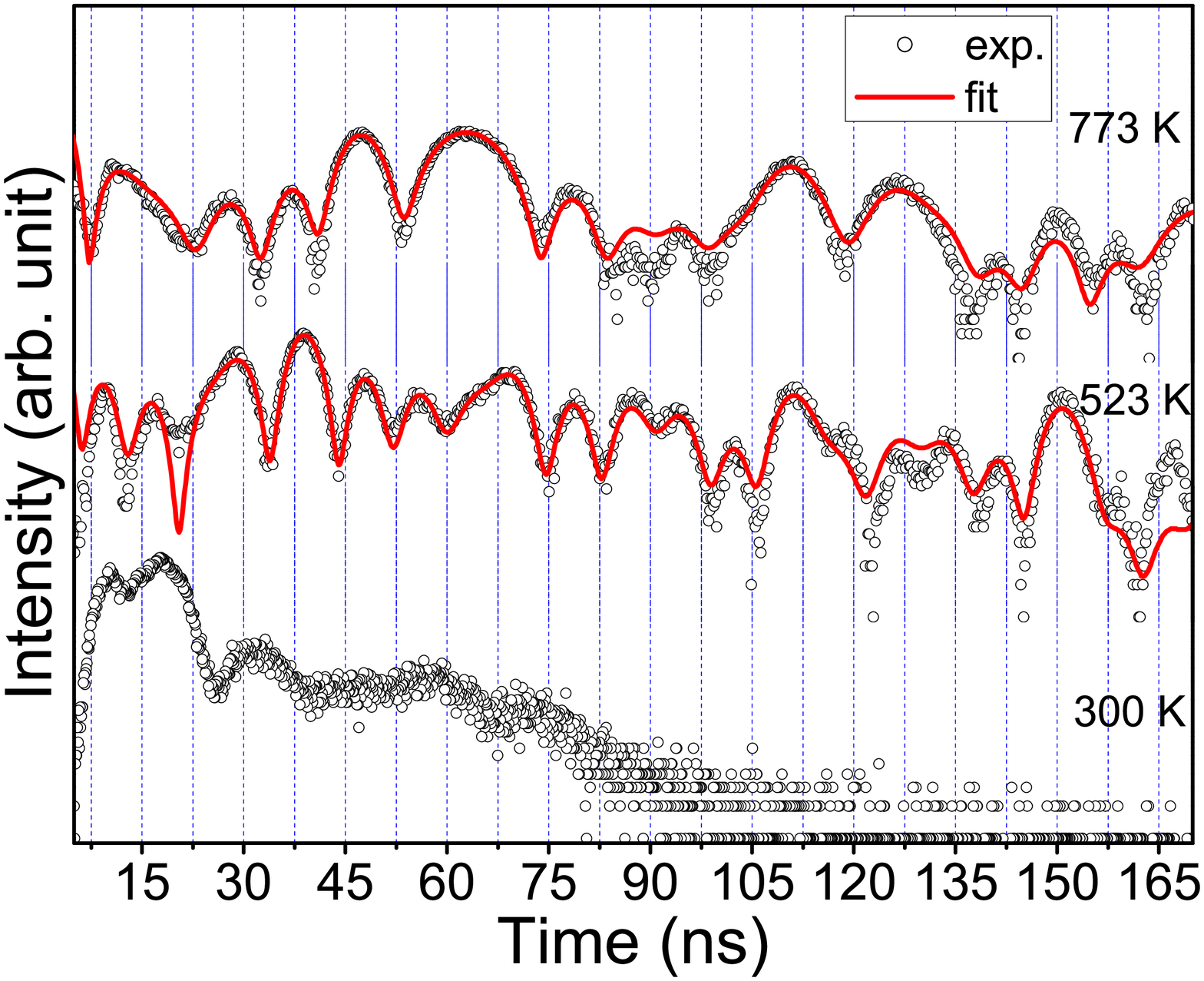} \vspace{-2mm}
\caption{\label{nrs} (Color online) Nuclear resonant scattering
spectra of the \FeC~samples deposited 300, 523 and 773 K measured
at grazing incidence of 0.21$^{\circ}$ at P01 PETRA III, Germany.}
\end{figure}

NRS is a fourier transform of M\"{o}ssbauer spectroscopy. This
technique is very sensitive to spatial phase factors due to
coherent scattering of radiation with matter. It gives a
possibility to correlate internal fields with the spatial
arrangement of the atoms~\cite{Rhlsberger2005NRS}. Now a days, the
availability of enormous brilliance of the synchrotron radiation
sources have made it possible to perform such kind of
measurements. This technique is frequently used to investigate
magnetic properties and phase transitions under high
pressure/temperature of nano-structure, ultrathin films,
clusters~\cite{PETRA3} and diffusion in the layered
systems~\cite{MG2005NRR}.

NRS spectra of of the \FeC~samples deposited at various \Ts~are
shown in the fig.\ref{nrs}. The spectrum of the sample deposited
at \Ts~= 300\,K shows few quantum beats (QBs) with a period of
about 15\,ns confirming the magnetic nature of this sample in
accordance with CEMS results~\cite{MIGLIERINI2017PAC}. But the NRS
signal decays soon after 40\,ns which can be understood due to
disordered local structure present in the amorphous
phase~\cite{MIGLIERINI2017PAC}. The NRS signal lasts much longer
times ($>$ 165\,ns) in samples deposited at higher \Ts. The QB
period of 523\,K sample varies from 5 to 15\,ns. On the other
hand, it varies from 10 to 15\,ns for the sample deposited at
773\,K. The smaller QB period reveals presence of a high magnetic
moment phase at 523\,K. To get more detail information, we fitted
NRS spectra of samples deposited at high \Ts~using REFTIM software
package~\cite{REFTIM2008}. The spectrum of the sample deposited at
\Ts~= 523\,K can be best fitted using combination of three
hyperfine fields, \bhf~= 21, 33, 34.3\,T. As already discussed,
\bhf~= 21 and 33\,T are respectively related to
$\mathrm{\theta}$-Fe$_{3}$C and $\mathrm{\alpha}$-Fe, the
additional component with larger \bhf~= 34.3\,T can be assigned to
Fe$_{4}$C~\cite{LIU2016SR} phase only. Their relative volume
fractions comes out to be 45\p~for $\theta$-Fe$_{3}$C, 35\p~for
$\alpha$-Fe and 20\p~for Fe$_{4}$C. On the other hand, the
spectrum of the sample deposited at \Ts~= 773\,K can be best
fitted assuming two components with \bhf~= 21 and 33\,T and their
relative volume fractions are 70 and 30\p, respectively. The
absence of \bhf~= 34.5\,T shows that higher \Ts~(773\,K) is not
favorable for the growth of Fe$_{4}$C phase.

\subsection{Depth profiling and Fe self-diffusion measurements}\label{3.4}

Fig.~\ref{sims}(a) and (b) shows SIMS depth-profiles of \FeC~and
pure Fe samples deposited at different \Ts. The sandwiched
$^{57}$Fe layer in both sets of sample results in a peak and it
becomes broadened with increase increase in \Ts. In \FeC~samples
$^{57}$Fe peak width ($\Delta_\mathrm{Fe}$) is about 4 and 8\,nm
at \Ts~= 300 and 523\,K, respectively and at \Ts~= 773\,K it has
completely diffused throughout the film. On the other hand,
$^{57}$Fe profile in pure Fe is significantly broader already at
300\,K but the broadening does not increase as much as in Fe-C at
higher \Ts. In addition, C concentration has also been estimated
from SIMS depth profiles (not shown) and it comes out to be 20, 17
and 13\pat~in samples deposited at \Ts~= 300, 523 and 773\,K,
respectively. This shows, C content decreases significantly for
the sample deposited at \Ts~= 773\,K. In accordance with XANES
spectra, decrease in C content with \Ts~shows that C is moving out
from the bulk of the sample.

As can be seen in Fig.~\ref{sims}(a), the trailing side of SIMS
profiles are broader than the rising side. Such broadening is
observed due to involvement of sputtering and small intermixing
produced by 3\,keV O$^{+}$ ions. Such profiles can be corrected
using following
equation~\cite{BREBEC1980ACTAMETAL,LOIRAT2000JNCS,MG2002PRB}:

\begin{equation}\label{E1}
    c_{c}(x+h)=c_{e}(x)+h\frac{dc_{e}(x)}{dx}
\end{equation}

where, \emph{c}$_{c}$ is corrected and \emph{c}$_{e}$ is
experimentally measured concentration profiles and $h$ is a
parameter representing the strength of intermixing. The value of
$h$ was kept constant for a series of samples.

To determine diffusion that is taking place during the growth of
our samples, the shape of the tracer profile can be represented as
a function of depth (\emph{x}) as:
\begin{equation}\label{E2}
    c(x,t)=\frac{c}{2\sqrt{2Dt}}\exp[-(x^{2}/4Dt]
\end{equation}

where, $c$ is a constant, \emph{t} is annealing time and \emph{D}
is diffusion coefficient. Therefore, profiles can be fitted using
a Gaussian function and diffusion coefficient can be calculated
using the following equation~\cite{LOIRAT2000JNCS}:
\begin{equation}\label{E3}
    D(t)=\frac{\sigma^{2}_{t}-\sigma^{2}_{0}}{2t}
\end{equation}

where, \emph{D(t)} is time average diffusion coefficient and
$\sigma_{t}$ is standard deviation of the Gaussian depth profile
over an annealing time of \emph{t} or when \emph{t} = 0.

\begin{figure} \center
\includegraphics [width=0.4\textwidth] {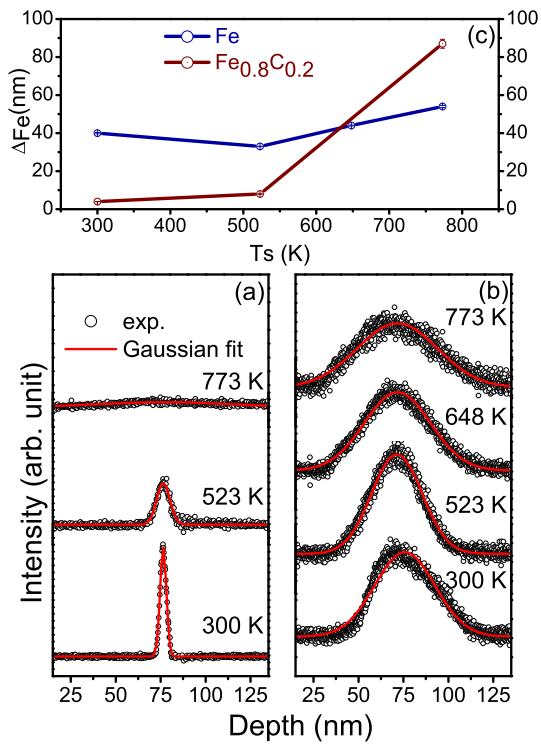} \vspace{-2mm}
\caption{\label{sims} (Color online) $^{57}$Fe SIMS concentration
profile of samples deposited at different substrate temperature
for \FeC~films (a) and that of pure Fe films (b). Obtained
$^{57}$Fe peak width
($\Delta_\mathrm{Fe}$) of these samples are compared in (c).} 
\end{figure}

Obtained $\Delta_\mathrm{Fe}$ in \FeC~and Fe samples are compared
in fig.~\ref{sims} (c). As can be seen that $\Delta_\mathrm{Fe}$
is about 10 times smaller in \FeC~as compared to Fe sample
deposited at \Ts~= 300\,K. Such a large variation in
$\Delta_\mathrm{Fe}$ is somewhat puzzling and unexpected. It is
known that fast grain boundary (\emph{gb}) diffusion takes place
in Fe due to defects or voids that are incorporated during the
growth. Addition of C seems to suppress them significantly.
Generally, it is anticipated that \emph{gb} diffusion would take
place at moderate temperatures. In our case, the information about
Fe self-diffusion during growth is obtained, it is new and unique
information and can be suitably used to understand the growth of
Fe based thin films and also C can be used as a effective dopant
to suppress Fe diffusion.

As we increase the~\Ts, $\Delta_\mathrm{Fe}$ increases albeit a
small drop in pure Fe deposited at \Ts~= 523\,K. Such a drop in
$\Delta_\mathrm{Fe}$ can be due to an interface sharpening effect
which happens due to release of defects and voids. Such interface
sharpening was also evidenced recently in Fe thin films grown at
573\,K~\cite{ATIWARI2019PRB} and also observed in earlier
works~\cite{BAI1996JPCM, KORTRIGHT1991JAP, Ishino2007JAP,
AMIR2012JAC}. At \Ts~=~523\,K, $\Delta_\mathrm{Fe}$ in \FeC~sample
is still significantly smaller as compared to Fe but when samples
were grown at \Ts~= 773\,K, a sudden rise in $\Delta_\mathrm{Fe}$
can be seen in \FeC~sample. It appears that at low \Ts~the
presence of C suppress Fe self-diffusion but when the
\Ts~increases beyond a particular value, Fe diffusion gets
augmented. Such kinetics of Fe self-diffusion affects formation of
Fe-C phases and will be discussed later. The schematic of
diffusion process is shown schematically in
fig.~\ref{diffschematic}.

\begin{figure} \center
\includegraphics [width=80mm,height=70mm] {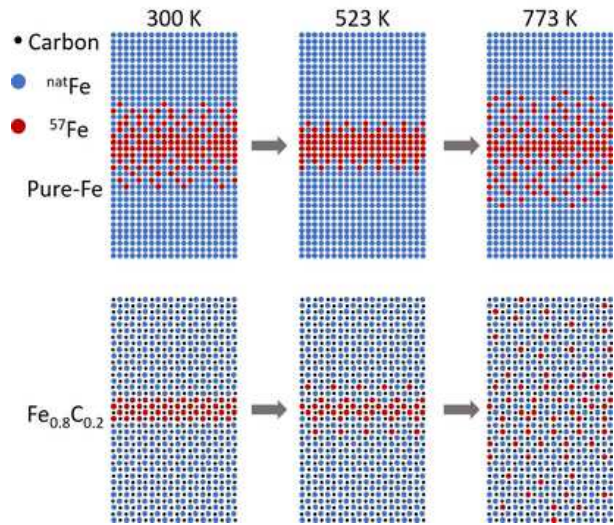} \vspace{-2mm}
\caption{\label{diffschematic} (Color online) Schematic
representation of diffusion process taking place during growth of
\FeC and Fe thin films deposited at \Ts~= 300, 523 and 773\,K.}
\vspace{-5mm}
\end{figure}

Using the values of $\Delta_\mathrm{Fe}$ in Fe-C and Fe samples
grown at \Ts~= 773\,K, we estimated Fe self-diffusion and it comes
out to be about an order of magnitude faster in \FeC~as compared
to Fe (6$\pm$1$\times$10$^{-19}$m$^{2}$/s in Fe-C and
7$\pm$3$\times$10$^{-20}$ m$^{2}$/s in Fe). Fe self-diffusion
coefficient obtained in our sample is close to the value found in
earlier works~\cite{LUBBEHUSEN1990AMM, Iijima2005JPED,
SUJAY2016IOP}.

\subsection{Phase transformation mechanism}\label{3.5}
From the results obtained in this work, a phase transformation
mechanism can be drawn to understand the formation of Fe-C phases
during the co-sputtering process. In co-sputtering process two or
more targets are sputtered simultaneously (here $\alpha$-Fe (bcc)
and graphite C targets). The mixing of sputtered Fe and C adatoms
takes place when they are still in the vapor phase. In the
sputtering process typically the adatom energy is about 10\,eV and
as adatom condense on a substrate they lose this energy in
picosecond time yielding quenching rates of the order of 10$^{16}$
K/s~\cite{WASA2012WAPSputtering}. These rates are about 10 orders
of magnitude higher as compared to melt roller
quenching~\cite{WASA2012WAPSputtering}. Generally, with such a
high quenching rates, the resulting phase should become amorphous
but this is certainly not the case as metallic samples produced by
sputtering do possess a long range ordering. This can be amply
seen from the XRD patterns of pure Fe films shown in
fig.~\ref{xrd}. Films grown at 300\,K are poly crystalline with an
average grain size of 18$\pm$0.5\,nm and with an increase in \Ts,
the grain size increases to 28$\pm$0.5\,nm at 523\,K,
45$\pm$1.5\,nm at 648\,K and 46$\pm$2\,nm at 773\,K. This clearly
indicates that after condensation on a substrate, the adatom
mobility driven diffusion process results in formation of long
range ordering and with an increase in \Ts, it increases. As such
this is trivial information which is well-known for growth of thin
films with sputtering~\cite{BHAVANASI2013OM, LI2009OpCm}. However,
this will be useful to understand the role of C in affecting phase
formation in Fe-C thin films.

Aforementioned, we placed a 3\,nm \ife~marker layer between
natural Fe layers, and through this we could measure Fe
self-diffusion that is taking place during the growth of film. We
found that already at 300\,K, the broadening in \ife~profile,
$\Delta_\mathrm{Fe}$ is quite large at about 40\,nm which is more
than ten times of the thickness of marker layer. However, with an
increase in \Ts~, $\Delta_\mathrm{Fe}$ does not increase as much.
Therefore, it appears that Fe self-diffusion takes place rapidly
during the initial stages of growth and thereafter it reduces
significantly. Fu \emph{et al}.~\cite{FU2005NATURE} did
multi-scale modelling of defect kinetics in iron and found that
the activation energy (E) for interstitial migration can be as low
as 0.3\,eV in $\alpha$-Fe. In an experimental study on Fe
self-diffusion in Fe/\ife~multilayers, it was also found E was
small (E$<<$1\,eV) and has been explained in terms of structural
defects in Fe that lead to fast Fe diffusion during initial stages
which subsequently becomes smaller when defects relaxation process
gets completed~\cite{SUJOY2009PRB}. In a recent study also, fast
Fe diffusion has been observed and explains in terms of triple
junctions leading to short-circuit
diffusion~\cite{ATIWARI2019PRB}. In a way, the fast Fe diffusion
during initial stages can be understood as grain boundary ($gb$)
diffusion. When the \emph{gb} diffusion gets over, annihilation of
defect causes Fe atoms to diffuse through a classical volume type
diffusion via thermal vacancies with very high E $\approx$ 3\,eV.

The addition of C in Fe affects the \emph{gb} diffusion process,
so much that $\Delta_\mathrm{Fe}$ $\rightarrow$ 0. Instead of
40\,nm for Fe, the $\Delta_\mathrm{Fe}$ in Fe-C was about 4\,nm,
close to its nominal thickness of 3\,nm. And within experimental
accuracy it can be inferred that $\Delta_\mathrm{Fe}$ $\approx$ 0
in Fe-C as compared to pure Fe. In this scenario, C atoms restrict
the path of Fe atoms thereby leading to formation of an amorphous
structure as observed in our Fe-C samples and also in previous
studies~\cite{BABONNEAU2000JAP, MI2004JPCM, FURLAN2015JPCM}.
However, when \Ts~increases to 773\,K even more rapid Fe diffusion
takes place, compared to the case when C was not added as shown in
fig.~\ref{sims} (c). Such an enhancement clearly indicates that in
presence of C, the concentration of defects may become even higher
leading to faster Fe diffusion through $gb$. But at an
intermediate temperature of 523\,K, we found that Fe diffusion was
still low and crystalline Fe-C phases like \tifc~and \tefc~start
to nucleate. And at this temperature regime, it seems that
kinetics of Fe-C phase formation is driven by C diffusion.
Recently, it has been revealed in a computational ReaxFF study
(based on bond order concept~\cite{DUIN2001REAXFFJPCA}) that C
diffuses through \emph{gb}~\cite{KUAN2018JPCC} and E for C
diffusion is typically about 0.8\,eV. Also, as suggested by
theoretical calculations, the energy barrier for $\alpha$
(bcc)~$\rightarrow$~$\gamma$ (fcc) phase transformation of Fe is
about 0.137\,eV/atom but it gets reduced to 0.127\,eV/atom in
presence of C~\cite{NGUYEN2018MT,NGUYEN2018CMS}. On the other
hand, for the reverse case i.e. $\gamma~\rightarrow~\alpha$ phase
it is much smaller at about 0.025\,eV/atom for Fe but it increases
marginally to 0.047\,eV/atom for Fe-C. The presence of C in Fe
lattice produces local stress field, resulting enhancement in the
energy barrier for
$\gamma~\rightarrow~\alpha$~\cite{NGUYEN2018MT,NGUYEN2018CMS}.
Therefore, the presence of C prevents $\gamma~\rightarrow~\alpha$
and favors the $\alpha~\rightarrow~\gamma$ phase transformation.
These conditions are suitably met at the intermediate temperature
of 523\,K and by further fine tuning the amount of C and
\Ts~around 523\,K, it may be possible increase the fraction of
\tefc~phase or even a single phase \tefc~phase can be obtained.

\section{Conclusion}
In conclusion, in the present work we systematically studied the
role of substrate temperature and phase formation in Fe-C thin
films around \FeC~composition. A comparison of \FeC~films together
with pure Fe films grown under similar conditions exhibited the
effect of C inclusion of on the long range crystalline ordering.
In addition, the comparison of \FeC~films with C thin films
yielded vital information about the hybridization between Fe and
C. By inserting a thin $^{57}$Fe or $^{57}$\FeC~marker layer in
between thick Fe or \FeC~layers, Fe self-diffusion that is taking
place during the growth itself was measured. We found the Fe
self-diffusion was appreciably large even at 300\,K, but the
addition of C in Fe inhibits Fe self-diffusion remarkably. At the
high \Ts~of 773\,K, C addition leads to very rapid Fe
self-diffusion. However, at an intermediate temperature of \Ts~ of
523\,K, Fe self-diffusion is still low and controllable so that
formation of \tefc~phase could be realized. It can be anticipated
that by further fine tuning of \Ts~and C composition, the fraction
of \tefc~can be further enhanced. The information about such Fe
diffusion process in Fe-C system is new and can be suitably used
to synthesize challenging Fe-C phases.

\section*{Acknowledgments}
Authors would like to acknowledge Layanta Behera for technical
help, Anil Gome for CEMS, Mohan Gangrade for AFM and MFM, Rakesh
Sah for XANES and Nidhi Pandey for NRS measurements. We are
thankful to A.\,K.\,Sinha for support and encouragements and Seema
for fruitful discussions. Portions of this research were carried
out at the light source PETRA III of DESY, a member of the
Helmholtz Association (HGF). Financial support by the Department
of Science $\&$ Technology (Government of India) provided with in
the framework of the India$@$DESY collaboration is gratefully
acknowledge.
\section*{References}

%

\end{document}